\documentclass[11pt]{article}
\usepackage{moriond,epsfig}
\usepackage{graphicx}
\usepackage{amsmath}
\usepackage{amssymb}
\usepackage{color}
\usepackage{graphicx}

\def\be{\begin{equation}}
\def\ee{\end{equation}}
\def\bea{\begin{eqnarray}}
\def\eea{\end{eqnarray}}

\newcommand{\GeV}{\,\mathrm{GeV}}

\newcommand{\Fig}[1]{Fig.~\ref{fig:#1}}

\newcommand{\eq}[1]{eq.~(\ref{eq:#1})}

\newcommand{\omi}[1]{}

\DeclareMathOperator{\im}{Im}
\DeclareMathOperator{\re}{Re}

\newlength{\myem}
\newcounter{mysubequation}[equation]

\newcommand{\MM}{\mathcal{M}^2}

\newcommand{\W}{\mathcal{W}}

\newcommand{\bsg}{B\to X_s\gamma}
\newcommand{\BRbsg}{\text{BR}(\bsg)}

\def\gsim{\lower.7ex\hbox{$\;\stackrel{\textstyle>}{\sim}\;$}}
\def\lsim{\lower.7ex\hbox{$\;\stackrel{\textstyle<}{\sim}\;$}}

\def\beq{\begin{equation}}
\def\eeq{\end{equation}}
\def\bea{\begin{eqnarray}}
\def\eea{\end{eqnarray}}

\begin{document}
\vspace*{4cm}
\title{Flavor Violation and Hierarchical  Sfermions}

\author{Marco Nardecchia}

\address{SISSA/ISAS and INFN, I--34014 Trieste, Italy}

\maketitle

\begin{abstract}
Naturalness arguments do not forbid the possibility that the first two families of squarks and sleptons are heavier than the rest of  the supersymmetric spectrum. In this framework, we study the phenomenology related to the flavor physics and we give bounds on the flavor violating parameters that we compare with the case of nearly degenerate squarks. The peculiar structure of the hierarchical scheme allows us to make definite predictions and suggests also a natural size for the flavor violating parameters.
\end{abstract}

\section{Framework of Hierarchical Sfermions}
The presence of the softly broken sector in the MSSM introduces a large number of new physical parameters. In particular, compared to the Standard Model (SM),  we have additional 36 mixing angles and 40 phases that give rise to flavor and CP violation.

The requirement that the supersymmetric and the SM contributions to physical observables agree with the  experimental data, gives strong constraints on the flavor-breaking structure of the soft terms in the MSSM.
In particular at least one of the following conditions is needed in order to suppress large supersymmetric contribution to a generic
 FCNC process:
\begin{itemize}
\item {\it Degeneracy.} The masses of  the sfermions present in the loop have almost the same values
\item {\it Alignment.}  The assumption is that quark and squark mass matrices are nearly simultaneously diagonalized by a supersymmetric field rotation, either in the down or in the up sector~\cite{nirs}.
\item {\it Irrelevancy.} The suppressions is obtained if the particles in the loop are very heavy.  
\end{itemize}

We study flavor physics in the framework of hierarchical soft terms, in which the first two generations of squarks and sleptons are heavier than the rest of the supersymmetric spectrum.
The flavor structure of the first and second generation squarks is tightly constrained by $K$ physics. On the other hand, the upper bounds on the masses of the first two generations of squarks are much looser than for the other supersymmetric particles. Therefore one can relax the flavor constraints, without compromising naturalness, by taking the first two generations of squarks much heavier than the third~\cite{dimg,ckn}. This procedure alleviates, but does not completely solve, the flavor problem and a further suppression mechanism for the first two generations must be present. However, it is not difficult to conceive the existence of such a mechanism which operates if, for instance, the soft terms respect an approximate U(2) symmetry acting on the first two generations~\cite{Pomarol:1995xc,u2}. In the case of hierarchy~\cite{Giudice:2008uk}, the small expansion parameter describing the flavor violation is the mismatch between the third-generation quarks identified by the Yukawa coupling and the third-generation squarks identified by the light eigenstates of the soft-term mass matrix. This small mismatch can be related to the hierarchy of scales present in the squark mass matrix and to CKM angles. However, for the phenomenological implications we are interested in, we do not have to specify any such relation and we can work in an effective theory where the first two generations of squarks have been integrated out. Their only remnant in the effective theory is the small mismatch between third-generation quarks and squarks.

\section{Hierarchy vs Degeneracy in Flavor Violating Amplitudes}
Let us consider the gluino contribution to a $\Delta F = 1$ process in the left-handed down quark sector, $d^L_i\rightarrow d^L_j$, neglecting for simplicity chirality changes. The amplitude of such a process is proportional to 
\begin{equation}
  \label{eq:ampl}
  A(\Delta F=1)\equiv f\left(\frac{\MM_D}{M_3^2}\right)_{d^L_id^L_j} =   \W_{d^L_i\tilde D_I}
  f\left(\frac{m^2_{\tilde D_I}}{M_3^2}\right) \W^*_{d^L_j \tilde D_I }.
\end{equation}
Here $f$ is a loop function, $M_3$ is the gluino mass and $\W$ is the unitary matrix diagonalizing the 6$\times$6 down squark squared mass matrix $\MM_D$ in a basis in which the down quark mass matrix is diagonal. We can simplify \eq{ampl} by using a perturbative expansion in the small off-diagonal entries of the squark mass matrix. 
The ``degenerate'' case is obtained in the limit in which the squark masses in the loop function coincide:
\begin{equation}
  \label{eq:DMI}
  f\left(\frac{\MM_D}{M_3^2}\right)_{d^L_id^L_j} = x f^{(1)}(x)  \, \delta^{LL}_{ij}, 
  \qquad
\text{(degenerate case)}
\end{equation}
where $x = \tilde m^2/M_3^2$ and $f^{(n)}$ is the $n$-th derivative of the function. The $\delta$ parameters are in this case normalized to the universal scalar mass $\tilde m^2$. 

In the ``hierarchical'' limit, the contribution to the loop function in \eq{ampl} from the heavy squarks is negligible. Therefore \eq{ampl} becomes
\begin{equation}
  \label{eq:HMI}
  f\left(\frac{\MM_D}{M_3^2}\right)_{d^L_id^L_j}= 
  f(x) \, \hat\delta^{LL}_{ij}. \qquad
 \text{(hierarchical case)}
\end{equation}
Here $x = \tilde m^2/M^2_3$ as before, where now  $\tilde m^2$ is interpreted as the third-generation squark mass. We have defined $\hat\delta^{LL}_{ij} \equiv \W_{d^L_i\tilde b_L} \W^*_{d^L_j \tilde b_L}$. Note that $\hat\delta^{LL}_{a3} \approx - (\MM_D)_{d^L_ad^L_3}/\tilde m^2_{a}$, so that $\hat\delta^{LL}_{a3}$ is again a normalized mass insertion. Also, $\hat\delta^{LL}_{12} = \hat\delta^{LL}_{13}(\hat\delta^{LL}_{23})^*$. 

For $\delta = \hat\delta$ the difference between the two schemes, the degenerate and the hierarchical one, is given by the order one difference between a function and its derivative. However, this difference becomes larger when we consider $\Delta F = 2$: 
\begin{equation}
  \label{eq:DHMI}
  A(\Delta F=2) =
    \begin{cases}
    \displaystyle
    \frac{x^2}{3!} g^{(3)}(x) (\delta^{LL}_{ij})^2 &
    \text{(degenerate case)} \\[3mm]
    g^{(1)}(x) (\hat\delta^{LL}_{ij})^2 &
    \text{(hierarchical case).}
  \end{cases}      
\end{equation}
Therefore, if $\tilde m^2$ is the same in the two cases we find that the amplitudes for $\Delta F=1$ and $\Delta F=2$ processes satisfy the relation
\begin{equation}
  \label{eq:2vs1}
  \left.\frac{A(\Delta F = 2)}{[A(\Delta F = 1)]^2}\right|_{\text{degenerate}} =
  \frac{g^{(3)}}{6g^{(1)}} \left(\frac{f}{f^{(1)}}\right)^2
  \left.\frac{A(\Delta F = 2)}{[A(\Delta F = 1)]^2}\right|_{\text{hierarchical}}.
\end{equation}
In general the ratio $(g^{(3)}/6g^{(1)})(f/f^{(1)})^2$ is typically small. As a consequence, the bounds on the $\Delta F=2$ processes inferred from $\Delta F=1$, or viceversa, may be significantly different in the two frameworks.

\section{Phenomenology of Hierarchical Sfermions}
The bounds on the flavor-violating parameters $\hat \delta$ are summarized in Table~\ref{tab:bounds} and compared with the bounds obtained in the case of degeneracy. An early analysis of the hierarchical case was presented in ref.~\cite{Cohen:1996sq}.
It is plausible to expect that the size of the parameters $\hat\delta^{LL}_{sb}$, $\hat\delta^{LL}_{db}$ cannot be smaller than the corresponding  CKM angles, $|V_{td}|$, $|V_{ts}|$ respectively \cite{Giudice:2008uk}. 
Thus, it is particularly interesting to probe experimentally flavor processes up to the level of $|\hat\delta^{LL}_{db}|\approx 8\times 10^{-3}$, $|\hat\delta^{LL}_{sb}|\approx 4\times 10^{-2}$ and $|\hat\delta^{LL}_{ds}|=|\hat\delta^{LL}_{db} \hat\delta^{LL*}_{sb} |\approx 3\times 10^{-4}$. The present constraints on the $b \leftrightarrow d$ transitions and on $\epsilon_K$ are at the edge of probing this region. An interesting conclusion is that hierarchical soft terms predict that new-physics effects in $b \leftrightarrow s$ transitions can be expected just beyond the present experimental sensitivity. 

\begin{table}
\begin{equation*}
\renewcommand{\arraystretch}{1.5}
\begin{array}{|c|c|c|}

\hline
D_0-\bar{D}_0  &
\left|  \hat \delta^{LL}_{ut} \hat \delta^{LL*}_{ct}\right| < 
8.0 \times 10^{-3} \left( \frac{m_{\tilde{t}}}{350 \ \textrm{GeV}} \right) &
\left|  \delta^{LL}_{uc} \right| < 
3.4 \times 10^{-2} \left( \frac{m_{\tilde{q}}}{350 \ \textrm{GeV}} \right) \\

\hline
B \to X_s \gamma &
\renewcommand{\arraystretch}{1.5}
\big| \re \big( \hat \delta^{LL}_{sb} \big) \big| < 
2.2 \times 10^{-2} \left( \frac{m_{\tilde{b}}}{350 \ \textrm{GeV}} \right)^2 \left( \frac{10}{\tan \beta} \right) &
\left| \re \left( \delta^{LL}_{sb} \right)  \right| < 
3.8 \times 10^{-2} \left( \frac{m_{\tilde{q}}}{350 \ \textrm{GeV}} \right)^2  \left( \frac{10}{\tan \beta} \right) \\
&
\big| \im \big( \hat \delta^{LL}_{sb} \big)  \big|  < 
6.7 \times 10^{-2} \left( \frac{m_{\tilde{b}}}{350 \ \textrm{GeV}} \right)^2  \left( \frac{10}{\tan \beta} \right) &
\left| \im \left(\delta^{LL}_{sb} \right)  \right|  < 
1.1 \times 10^{-1} \left( \frac{m_{\tilde{q}}}{350 \ \textrm{GeV}} \right)^2  \left( \frac{10}{\tan \beta} \right) \\

\hline
\Delta m_{B_s} &
\renewcommand{\arraystretch}{1.5}
\big| \re \big( \hat \delta^{LL}_{sb} \big)  \big|  < 9.4 \times 10^{-2}  \left( \frac{m_{\tilde{b}}}{350 \ \textrm{GeV}} \right) &
\left| \re \left( \delta^{LL}_{sb} \right)   \right| <  4.0 \times 10^{-1} \left( \frac{m_{\tilde{q}}}{350 \ \textrm{GeV}} \right) \\
&
\big| \im \big( \hat \delta^{LL}_{sb} \big)  \big| < 7.2 \times 10^{-2} \left( \frac{m_{\tilde{b}}}{350 \ \textrm{GeV}} \right) &
\left| \im \left( \delta^{LL}_{sb} \right) \right|  <  3.1 \times 10^{-1} \left( \frac{m_{\tilde{q}}}{350 \ \textrm{GeV}} \right) \\

\hline
B_d^0\text{--}\bar B_d^0 &
\renewcommand{\arraystretch}{1.5}
\big| \re \big( \hat \delta^{LL}_{db} \big)  \big|  < 4.3 \times 10^{-3} \left( \frac{m_{\tilde{b}}}{350 \ \textrm{GeV}} \right)&
\left| \re \left( \delta^{LL}_{db} \right)  \right|  < 1.8 \times 10^{-2} \left( \frac{m_{\tilde{q}}}{350 \ \textrm{GeV}} \right) \\
&
\big| \im \big( \hat \delta^{LL}_{db} \big)  \big|  < 7.3 \times 10^{-3} \left( \frac{m_{\tilde{b}}}{350 \ \textrm{GeV}} \right) &
\left| \im \left(\delta^{LL}_{db} \right)   \right|  < 3.1 \times 10^{-2} \left( \frac{m_{\tilde{q}}}{350 \ \textrm{GeV}} \right) \\

\hline
\Delta m_{K} &
\renewcommand{\arraystretch}{2}
\sqrt{ \Big| \re \big( \hat{\delta}^{LL}_{db}  \hat{\delta}^{LL*}_{sb} \big)^2 \Big|} <  1.0 \times 10^{-2} \left( \frac{m_{\tilde{b}}}{350 \ \textrm{GeV}} \right) &
\sqrt{ \left| \re \left( \delta^{LL}_{ds} \right)^2 \right|} <   4.2 \times 10^{-2} \left( \frac{m_{\tilde{q}}}{350 \ \textrm{GeV}} \right)  \\

\hline
\epsilon_{K} &
\renewcommand{\arraystretch}{2}
\sqrt{ \Big| \im \big(  \hat{\delta}^{LL}_{db}  \hat{\delta}^{LL*}_{sb}  \big)^2 \Big|} < 4.4 \times 10^{-4} \left( \frac{m_{\tilde{b}}}{350 \ \textrm{GeV}} \right) &
\sqrt{ \left| \im \left( \delta^{LL}_{ds} \right)^2 \right|} < 1.8 \times 10^{-3}  \left( \frac{m_{\tilde{q}}}{350 \ \textrm{GeV}} \right)  \\

\hline
\end{array}
\end{equation*}
\caption{Bounds on the LL insertions in the hierarchical and degenerate cases. The limits on the RR insertions are the same, except the one from $\BRbsg$, which is much weaker. }
\label{tab:bounds}
\end{table}

\begin{figure}
\begin{center}
\includegraphics[width=\textwidth]{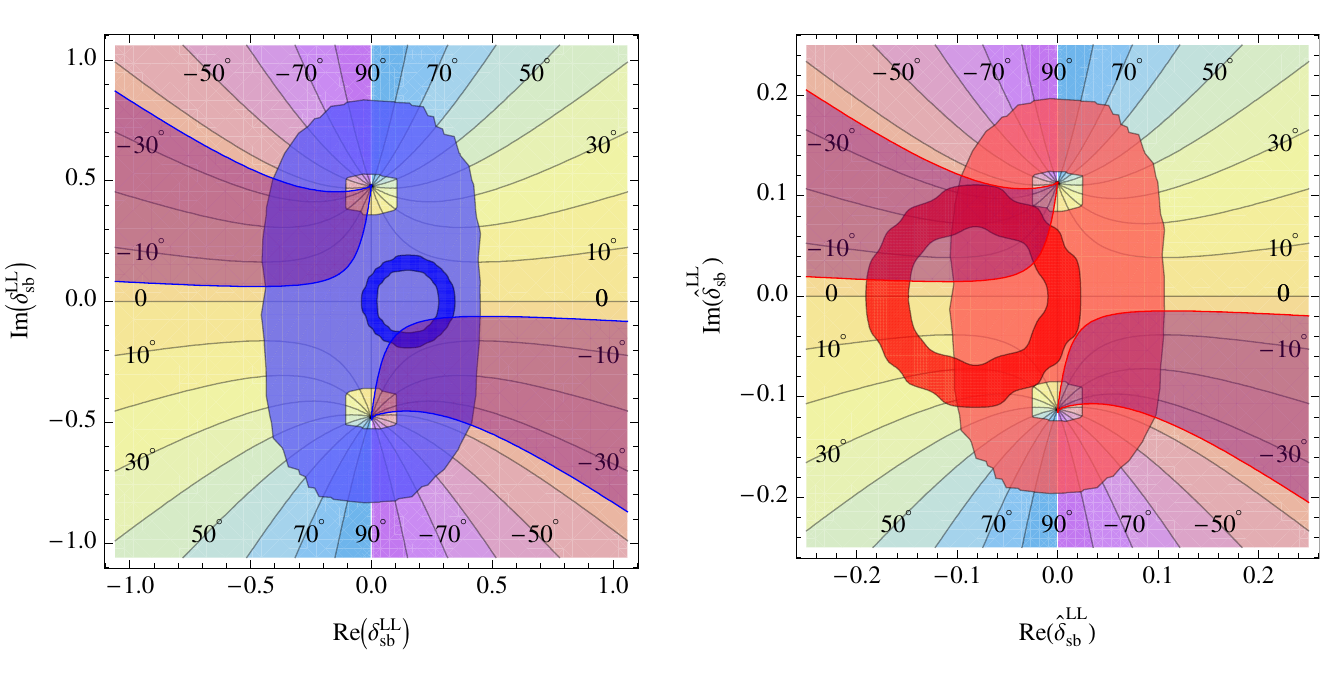}
\end{center}
\caption{95\% CL bounds on the real and imaginary parts of $\delta^{LL}_{sb}$ (left, blue) and $\hat\delta^{LL}_{sb}$ (right, red) from the measurements of $\Delta m_{B_s}$ (lighter shading) and $\BRbsg$ (darker shading) for $\tilde m = M_3  = \mu = 350\GeV$ and $\tan\beta =10$. In the background, the contour lines of the phase $\phi_{B_s}$ are shown.}
\label{fig:bs}
\end{figure}

Another interesting point regarding the phenomenology of the hierarchical framework is the fact that  the new-physics effects in $b\leftrightarrow s$ transitions are particularly promising.
For example recent measurements from the CDF~\cite{Aaltonen:2007he} and D0~\cite{:2008fj} collaborations have shown a mild tension between the experimental value and the SM prediction for the $\phi_{B_s}$  mixing phase, at the $2.5\,\sigma$ level~\cite{hfag}.
The hierarchical case allows values of the phase $\phi_{B_s}$ about three times larger than in the degenerate case, in agreement with the generic expectation from \eq{2vs1}. The range of $\phi_{B_s}$ presently favored by the experiment is shown in \Fig{bs}. 

\section*{Acknowledgments}
I thank Andrea Romanino and Gian Giudice for all the physics I learned from them, and the organizers for the opportunity to come and speak at Moriond EW09.

\end{document}